# Symmetry breaking in vanadium trihalides

Luigi Camerano[1] and Gianni Profeta[1,2]

[1]*Department of Physical and Chemical Sciences, University of L'Aquila, Via Vetoio, 67100 L'Aquila, Italy*
[2]*CNR-SPIN L'Aquila, Via Vetoio, 67100 L'Aquila, Italy*

In the light of new experimental evidence we study the insulating ground state of the $3d^2$-transition metal trihalides VX$_3$ (X=Cl, I). Based on Density Functional Theory with the Hubbard correction (DFT+$U$) we systematically show how these systems host multiple metastable states characterized by different orbital ordering and electronic behaviour. Our calculations reveal the importance of imposing a precondition in the on site $d$ density matrix and of considering a symmetry broken unit cell to correctly take into account the correlation effects in a mean field framework. Furthermore we ultimately found a ground state with the $a_{1g}$ orbital occupied in a distorted VX$_6$ octahedra driven by an optical phonon mode.

## I. INTRODUCTION

The use of Hubbard-like correction to the Density Functional Theory (DFT+$U$) is mainly motivated to properly take into account the localized nature of the $d$ and $f$ orbitals and to correctly describe the observed insulating behaviour in some strongly correlated materials [1–4]. In practical implementation of the method [5–8], the difficulty to precisely account for the localization of electrons can numerically produce convergence to different electronic metastable phases, depending on the initial charge density used in the calculations, like in symptomatic compounds such as FeO [1] and UO$_2$ [9]. These metastable phases could have very different energies and can constitute a trap preventing to access the *real* ground state of the system. Moreover, the stabilization of metastable electronic states in strongly correlated systems is sometimes due to what A. Zunger calls "simplistic" electronic structure theory: a mean-field calculation (like DFT in the local density approximation) in a unit cell showing as many symmetry constrains as possible [10, 11]. Exceptional confirmation of the theory are the predictions of insulating phases in "Mott insulators" even by mean-field like DFT providing different symmetry breaking (spin symmetry breaking, structural symmetry breaking, consideration of spin-orbit coupling (SOC)) [11] which allows to explore strong correlations and multireference character in the real symmetry unbroken wavefunction [12–14].

In this paper, we provide a further and unnoticed example of symmetry breaking induced phases in two dimensional magnetic materials, discovering that symmetry breaking is necessary to stabilize the right insulating ground state among different metastable ones hosted by layered magnetic VI$_3$ [15] and VCl$_3$ [16] compounds. We find that depending on the strength of on-site Coulomb repulsion (U), SOC and structural distortions, different orbital ordered phases can be stabilized.

The delicate balancing of these low energy interactions is ultimately originated by the common structural motif in VX$_3$ compounds: honeycomb lattice of V cations in edge sharing octahedral coordination with the halides [15–18]

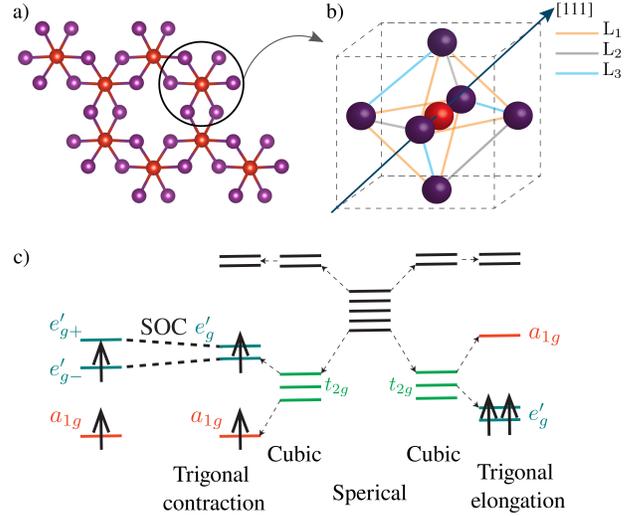

FIG. 1. a) Top view of the VX$_3$ monolayer: honeycomb lattice of V cations (red spheres) in edge sharing octahedral coordination with the halides X=Cl, Br, I (purple spheres). b) Octahedra in trigonal distorted phase ($D_{3d}$). In metal trihalides $L_1$, $L_2$, $L_3$ distances are inequivalent. The [111] direction is the trigonal axis and correspond to the $z$ out-of-plane direction. c) Schematic representation of orbital symmetry breaking under structural distortion. Starting from the spherical symmetry, we report $d$-level splitting in cubic symmetry and under trigonal distortion. Elongation and contraction refer to the trigonal axis. The effect of the spin-orbit coupling is also reported: the $e'_g$ orbitals split because of diffent angular momentum.

(Fig. 1a-b). The sixfold coordination of the metal cations could naturally lead to an octahedral symmetry O$_h$, but, due to partial occupation of $d$ orbital in $t_{2g}$ shell (vanadium is in a V$^{3+}$ oxidation state with two electrons in the $d$ correlated manifold), a Jahn-Teller distortion of the octhaedra lowers the symmetry from O$_h$ to trigonal point group D$_{3d}$ (Fig. 1b) [19, 20], determining the splitting of the $d$-state manifold according to the trigonal basis (see Fig. 1c and Table S1 in Supplementary Material (SM)). Similar octhaedral environment and structural distortion are commonly observed in 3$d$ ABO$_3$ perovskites [21–24].

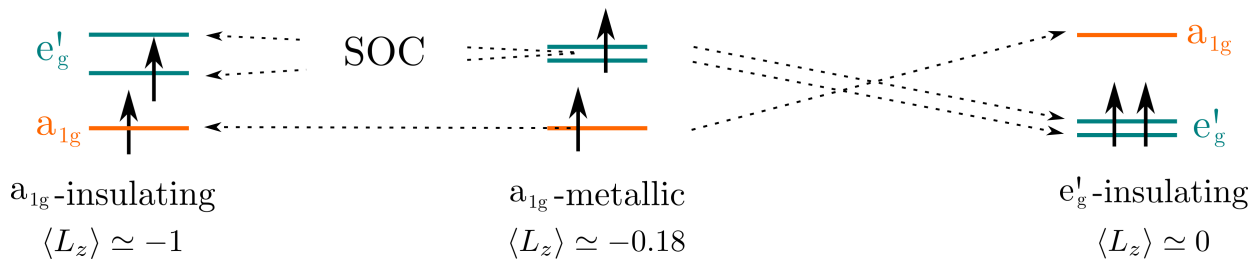

FIG. 2. Sketch of the possible phases that can be established in VX$_3$ compounds. Central panel: the a$_{1g}$-metallic phase with a partially quenched angular momentum and hybridized e'$_g$ orbitals. Left panel: the a$_{1g}$-insulating phase with an unquenched angular momentum. Right panel: the e'$_g$-insulating phase with a quenched angular momentum. The corresponding density of states and band structure are reported in the SM in Figs. S2,S3

As a general rule, the trigonal symmetry determines the splitting of $t_{2g}$ orbitals in a doublet $e'_g$ ($e'_{g-}$ and $e'_{g+}$) and a singlet $a_{1g}$ (Fig. 1c). The $a_{1g}$ orbital extends along the out-of-plane direction (see Table S1 in SM) and in these compounds corresponds to the cubic harmonic with $|l^z = 0\rangle$ [19, 25]. On the other hand, the $e'_{g-}$ and $e'_{g+}$ orbitals belong to the eigenspace with $|l^z = -1\rangle$ and $|l^z = 1\rangle$, respectively.

Layered VX$_3$ compounds are deeply investigated motivated by the technological perspectives they promise and for the peculiar physical properties they have demonstrated to host [16, 26, 27]. They have been recently investigated by different experimental and theoretical approaches because of contrasting evidences on the nature of their electronic ground state [16, 25, 28–31].

Transmission optical spectroscopy [32] and photoemission experiments [16, 29, 30, 33, 34] revealed a sizable band gap for both VI$_3$ and VCl$_3$. Polarization dependent angular resolved photoemission experiments [30] on VI$_3$ showed evidences that in-plane ($e'_g$ manifold) and out-of-plane ($a_{1g}$) orbitals are both occupied, pointing to a $a_{1g}e'_{g-}$ ground state, see Fig. 1c (here and after, we refer to this phase as $a_{1g}$-insulating ground state, see Fig. S1 in SM for additional details). Indeed, X-ray magnetic circular dichroism measurement [25] confirmed the presence of a large orbital moment ($\langle L_z\rangle_{exp} \simeq -0.6$), compatible with the $a_{1g}$-insulating ground state, although partially quenched with respect to the theoretically predicted $\langle L_z\rangle_{th} \simeq -1$ [25, 28].

Despite all these experimental evidences point to an $a_{1g}$-insulating ground state, recent photoemission spectra on VI$_3$ and VCl$_3$ [29, 30, 34] proved to be particularly difficult to interpret by first-principles DFT+$U$ calculation, in particular for the determination of the energy position of V-$d$ states [25, 29, 32, 35, 36]. To further complicate the scenario, different first-principle calculations [37, 38] predicted both VCl$_3$ and VI$_3$ to be metals. The emergence of false metallic states in first-principles DFT calculations [10] could be an indication of possible metastable electronic phases which prevent from obtaining the ground state.

In addition, the above cited difficulties to reconcile DFT+$U$ predictions with angular resolved photoemission spectroscopy (ARPES) experiments and the discrepancies between the experimental orbital angular momentum ($\langle L_z\rangle_{exp} \simeq -0.6$) and the theoretical prediction ($\langle L_z\rangle_{exp} \simeq -1$) call for a deeper understanding of the ground state properties of Vanadium based trihalides, solving both the numerical and physical aspects which hinder the determination of the electronic configuration in these compounds.

## II. METASTABLE PHASES

The electronic structure of VI$_3$ and VCl$_3$ are studied using a monolayer unit cell in the $D_{3d}$ point-group using DFT+$U$ approach including SOC and considering the spin polarization along the $z$-axis (see Methods section in SM for further information). As anticipated, DFT calculation in this class of materials is complicated by competing and entangled structural, electronic and magnetic degree of freedom, in which the self-consistent solution of the Kohn-Sham equation could be trapped depending on the starting guess for the charge density and wavefunctions. So, in order to stabilize different metastable states, we used a precondition on the onsite occupation of the $d$-density matrix for the $+U$ functional [39] (see Fig. S1 in SM for a summary on the $d$-states representation in a $D_{3d}$ structure and on the precondition matrices used in the calculations).

We start the discussion presenting the electronic phase we have obtained in VI$_3$ forcing the occupation of the $a_{1g}$-metallic state. The self-consistency, without SOC, ends in a phase with orbital angular momentum directed along the $z$ direction ($\langle L_z\rangle \simeq -0.18$) and the $a_{1g}$ orbital occupied (central panel of Fig. 2). In this configuration, VI$_3$ results metallic due to the hybridization of the $e'_g$ doublets with iodine derived bands (See Figs. S2, S3). We refer to this last phase as the $a_{1g}$-metallic phase (see SM for a better description of the technicalities used to stabilized the different phases).

On the other hand, forcing the occupation of the $a_{1g}$ state and including SOC in the calculation produces an insulating phase with occupied $a_{1g}$ state and with an orbital angular momentum $\langle L_z\rangle \simeq -1$ compatible with the occupation of the $e'_{g-}$ orbital (leftmost panel in Fig. 2). Finally, leaving the $a_{1g}$ empty in the starting guess for the density matrix, we find an insulating solution for both VI$_3$ and VCl$_3$ (rightmost panel in Fig. 2). In the insulating solution the $e'_g$ doublet is fully occupied and, as expected, the angular momentum is quenched $\langle L_z \rangle \simeq 0$. Indeed, Geourgescu et al. [20] claim that, in light compounds without strong SOC, such as VCl$_3$ and TiCl$_2$, the only insulating ground state must be the $e'_g$-insulating phase, driven by electronic-correlation (Hubbard U Coulomb repulsion). We note that the $a_{1g}$- and $e'_g$-insulating phases differ also from a structural point of view [19, 25]: the $a_{1g}$-insulating phase is trigonally contracted while the $e'_g$-insulating phase is trigonally elongated (see Fig. 1c).

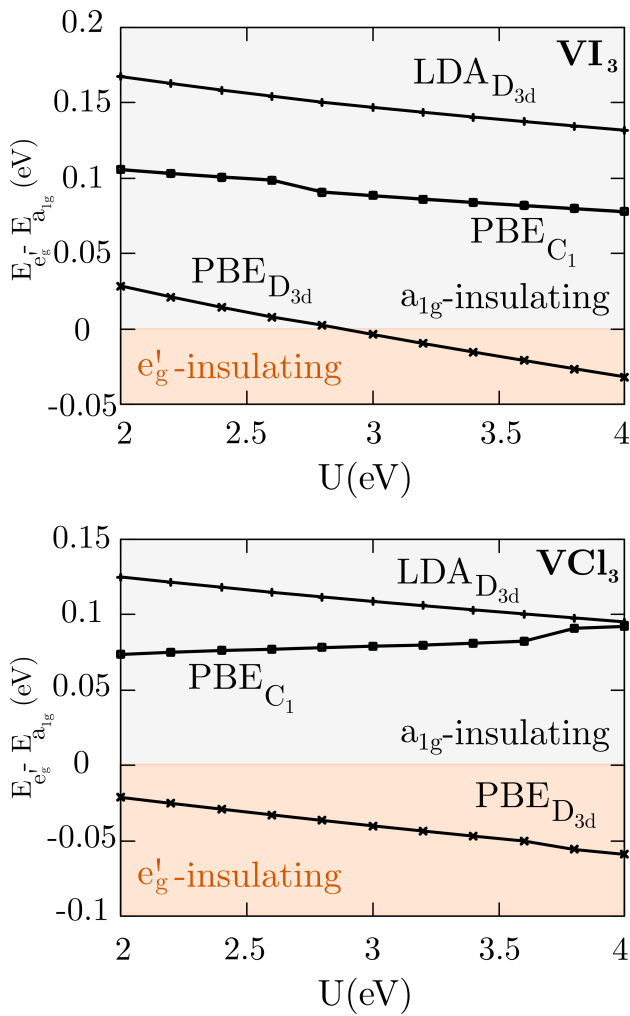

FIG. 3. Energy differences between $e'_g$-insulating phase and $a_{1g}$-insulating phase using PBE and LDA functional in a $D_{3d}$ unit cell. The $PBE_{C1}$ line represent the differences between the $e'_g$-insulating phase and the symmetry-broken (SB) $a_{1g}$-insulating phase in a unit cell with randomized positions of the atoms (see text).

A deeper analysis of the self-consistent $d$-density matrix reveals that both $a_{1g}$-metallic and $e'_g$-insulating phases have a block diagonal form with real occupancies, in agreement with the symmetry constrains, while the $a_{1g}$-insulating phase show a $d$-density matrix with imaginary occupancies signaling the fundamental role of SOC in stabilizing this state (see Fig. S1 in SM).

### III. CORRELATION EFFECTS

All the investigated phases lie close in energy (order of $\simeq$ 10-50 meV) and, given the already discussed delicate physical properties of this class of compounds, it is worth understanding how much the ground state properties could depend on the fine details of the calculation. Thus, we analyze the energy differences among the discovered phases as a function of U parameter and for two choices of the exchange-correlation potential, PBE and LDA. We present the results in Fig. 3a-b (curves labelled as PBE$_{D3d}$ and LDA$_{D3d}$) for VI$_3$ and VCl$_3$ respectively, where the total energy difference between the two insulating phases ($e'_g$ and $a_{1g}$) is reported as a function of the U parameter. [40]. We find that the total energy difference is strongly dependent on U for VI$_3$. In particular, for U less than $\simeq$ 3.0 eV the $e'_g$ is lower in energy, while the $a_{1g}$ is the ground state for larger U. For VCl$_3$, the $e'_g$ is always favored. On the other hand, the LDA functional predicts the $a_{1g}$-insulating phase as the ground state for both VCl$_3$ and VI$_3$, irrespective on U value. Note that, even in light compound, like VCl$_3$, SOC makes accessible the $a_{1g}$-insulating phase (that is the ground state in LDA approximation).

### IV. STRUCTURAL SYMMETRY BREAKING

The highlighted strong dependence of the total energies of the considered phases on both U and exchange-correlation potential point towards a complicated and nearly degenerate electronic energy landscape promoted by strong electronic correlations, which can be further enriched by the coupling with structural degree of freedom. Indeed, following the approach of Zunger et al. [10, 11] strong correlations in the exact wavefunction can be captured in a DFT+$U$ framework, lowering the symmetry inducing structural distortions. To explore this possibility in an unbiased way, we randomize all the positions of the atoms in the unit cell and then relaxed the system toward the closest local energy minimum. The calculation ends in a new, completely symmetry-broken (SB) phase [41] which we call PBE$_{C1}$ characterized by a distortion of the halides ($\sim$ 0.03 Å). We find that this last phase has an occupied $a_{1g}$ orbital and a nearly quenched angular momentum $\langle L_z \rangle \simeq -0.1$ and is energetically favored for all the U values studied (Fig 3a-b). This result is reminiscent of a sort of Jahn-Teller effect (similar of those found in perovskites [21–24]), in which structural distortion promote an electronic energy lowering. The effect can be further rationalized studying the phonon modes

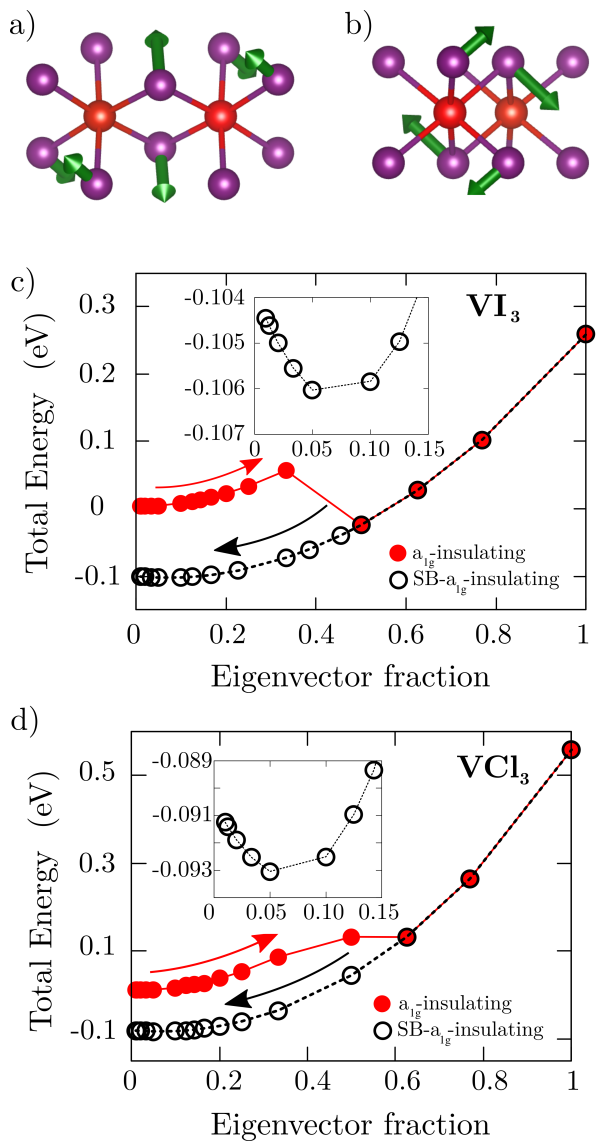

FIG. 4. Top (panel a) and side (panel b) view of the optical phonon mode at Γ. VI$_3$ (panel c) and VCl$_3$ (panel d) total energy as a function of the normalized eigenvector fraction. The filled red dot are energies obtained starting from the charge of $a_{1g}$-insulating phase that has imaginary occupancies in the $d$-density matrix, the empty black dots are energies obtained starting from a SB $a_{1g}$-insulating phase with real occupancies in the $d$-density matrix. The red dots with the black border are obtained starting from the $a_{1g}$-insulating phase but converging in a SB $a_{1g}$-insulating phase. In the insets there is a zoom close to the undistorted phase

at Γ-point, which could signal structural instabilities by imaginary frequencies. We found that the $a_{1g}$-insulating phase [42] have real frequencies, indicating that it is, indeed, a real *metastable* phase, stable for relatively small structural perturbations around the equilibrium structure. However, the calculation of the total energy displacing the atoms along the eigenvector of one optical phonon mode (mainly involving the halides as is shown in Fig. 4a-b) using larger distortions (up to $\simeq 0.2$ Å), reveals an electronic instability when the displacement of the atoms is 0.5 times the normalized eigenvector for VI$_3$ and $\sim 0.6$ for VCl$_3$ (see Figure 4c-d). Above these thresholds the systems do not remain in the same Born-Oppenheimer energy surface (red curve) but moves to a different one (black curve). Moving back along the eigenvector direction from this new electronic phase to recover the undistorted unit cell, both systems end in a new energy minimum clearly highlighted in the insets of Figure 4c-d, with inequivalent distances between V and X (of about 0.03Å) corresponding to an $a_{1g}$-insulating state with $\langle L_z \rangle \simeq -0.1$ and a $d$-density matrix with real occupancies, in agreement with the evidences from the most recent photoemission experiments [29, 30, 34] in VI$_3$. It must be underlined that this last phase cannot be obtained in DFT calculation in a fully symmetric $D_{3d}$ unit cell, because it sets a univocal orbital scheme (see Fig. 2). Moreover, it is not even stabilized by SOC: a calculation without SOC on this new phase, starting from the previous converged charge density remains equally stable.

## V. CONCLUSION

The newly discovered phase in Vanadium trihalides originates from the $a_{1g}$-metallic phase upon a Jahn-Teller distortion which splits the degenerate $e'_g$ doublet opening an energy gap further increased by electronic correlations, possibly signaling strong correlation in the true correlated wavefunction [10–13].
Our results shed light on new physical mechanisms active in Vanadium based trihalides, overlooked and unexplored so far, which can be theoretically accessed once the electronic correlations are considered in a structural SB unit cell. We can now interpret the available experimental reports considering the above predicted SB phase. X-ray natural linear dichroism and X-ray magnetic circular dichroism experiment by Sant *et al.* [31] found an anisotropic charge density distribution around the V$^{3+}$ ion which we could naturally intepret with an unbalanced hybridization between the vanadium and the ligand states, originating from the structural distortion and inequivalence of the ligands. The occupation of the $a_{1g}$ orbital and the insulating behavior, are confirmed by polarization dependent ARPES [30], but the predicted orbital moment ($\langle L_z \rangle \simeq -0.1$) results too far from the experimental estimation of $\langle L_z \rangle \simeq -0.6$ [25]. However, it should be noted that neither the symmetric $a_{1g}$-insulating ($\langle L_z \rangle \simeq -1$) nor the $e'_g$-insulating phase ($\langle L_z \rangle \simeq 0$) can explain the measured value, leaving this aspect open for further theoretical and experimental investigation.
We conclude calling for dedicated experiments to confirm the newly discovered SB phase and to possibly access the different metastable phases we discovered. In addition, the numerical technique based on pre-condition of the $d$- density matrix and the exploration of symmetry-broken phases can represent a valuable computational tool to study low energy phases of correlated magnetic systems.

## VI. DATA AVAILABILITY STATEMENT

All data that support the findings of this study are included within the article and supplementary materials.

## VII. ACKNOWLEDGEMENTS

We thank D. Mastrippolito for carefully reading the manuscript and valuable suggestions. G.P. acknowledges support from CINECA Supercomputing Center through the ISCRA project and financial support from the Italian Ministry for Research and Education through the PRIN-2017 project "Tuning and understanding Quantum phases in 2D materials-Quantum 2D" (IT-MIUR Grant No. 2017Z8TS5B). This work was funded by the European Union-NextGenerationEU under the Italian Ministry of University and Research (MUR) National Innovation Ecosystem Grant No. ECS00000041 VITALITY-CUP E13C22001060006.

Supplemental material and supporting information for

# The role of symmetry breaking in Vanadium trihalides

Luigi Camerano,[a], Gianni Profeta[a,b]

[a] Department of Physical and Chemical Sciences, University of L'Aquila, Via Vetoio 67100 L'Aquila, Italy
[b] CNR-SPIN L'Aquila, Via Vetoio 67100 L'Aquila, Italy

## CONTENTS

## I. METHODS

Density functional theory calculations were performed using the Vienna ab-initio Simulation Package (VASP) [1, 2], using both the generalized gradient approximation (GGA), in the Perdew-Burke-Ernzerhof (PBE) parametrization for the exchange-correlation functional [3] and local density approximation (LDA). Interactions between electrons and nuclei were described using the projector-augmented wave method. Energy thresholds for the self-consistent calculation was set to $10^{-5}$ eV and force threshold for geometry optimization $10^{-4}$ eV Å$^{-1}$. A plane-wave kinetic energy cutoff of 450 eV was employed for both VI$_3$ and VCl$_3$. The Brillouin zone was sampled using a $12 \times 12 \times 1$ Gamma-centered Monkhorst-Pack grid. To account for the on-site electron-electron correlation we used the GGA+U and LDA+U approaches with an effective Hubbard term $U = 3.5$ eV for VCl$_3$ and $U = 3.7$ eV for VI$_3$ consistent with the value calculated by He *et al.* [4] with linear response theory [5]. The VCl$_3$ and VI$_3$ monolayer phases were described using a lattice parameter of $a = 6.084$ [6] Å and $a = 6.93$ Å [7] respectively and a vacuum region of 15 Å. Phonons at the $\Gamma$-point have calculated by finite differences method [8].

## II. $d$-DENSITY MATRIX PRECONDITION

In Fig. S1 we report an overview of the different phases described in the manuscript (name of the phase, orbital angular momentum, stabilization method, $d$-density matrix occupation).

| | Phase | |
|---|---|---|
| a$_{1g}$-insulating | a$_{1g}$-metallic | e$'_g$-insulating |
| | **Orbital angular momentum** | |
| $\langle L_z \rangle \simeq -1$ | $\langle L_z \rangle \simeq -0.18$ | $\langle L_z \rangle \simeq 0$ |
| | **Phase stabilization** | |
| DFT+U calculation with SOC and forcing the occupation of the a$_{1g}$ orbital. | DFT+U calculation without SOC and forcing the occupation of the a$_{1g}$ orbital. Once converged, the inclusion of SOC does not modify the stabilized phase. | DFT+U calculation without SOC and starting from atomic orbitals. Once converged, the inclusion of SOC does not modify the stabilized phase. |
| | **$d$-density matrix** | |
| $\begin{bmatrix} 0.46 & 0.1 & 0 & -0.17i & 0.29i \\ 0.1 & 0.33 & 0 & -0.09i & 0.17i \\ 0 & 0 & 0.89 & 0 & 0 \\ 0.17i & 0.09i & 0 & 0.33 & -0.1 \\ -0.29i & -0.17i & 0 & -0.1 & 0.46 \end{bmatrix}$ | $\begin{bmatrix} 0.57 & 0.22 & 0 & 0 & 0 \\ 0.22 & 0.33 & 0 & 0 & 0 \\ 0 & 0 & 0.88 & 0 & 0 \\ 0 & 0 & 0 & 0.33 & -0.22 \\ 0 & 0 & 0 & -0.22 & 0.57 \end{bmatrix}$ $d^{\uparrow}_{xy}\ d^{\uparrow}_{yz}\ d^{\uparrow}_{z^2}\ d^{\uparrow}_{xz}\ d^{\uparrow}_{x^2-y^2}$ | $\begin{bmatrix} 0.61 & 0.30 & 0 & 0 & 0 \\ 0.30 & 0.57 & 0 & 0 & 0 \\ 0 & 0 & 0.08 & 0 & 0 \\ 0 & 0 & 0 & 0.57 & -0.30 \\ 0 & 0 & 0 & -0.30 & 0.61 \end{bmatrix}$ |

FIG. S1. Overview of the metastable phases in VX$_3$ compounds

In a $D_{3d}$ symmetry the $d$ orbitals can be written, with the $z$ axis directed along trigonal axis, as is shown in Table S1. In this representation the $d$-density matrix [9] is block diagonal.

| Cubic basis | | Trigonal Basis | |
|---|---|---|---|
| Symmetry | Orbitals | Symmetry | Orbitals |
| $t_{2g}$ | $d_{\bar{x}\bar{y}}$ | $a_{1g}$ | $d_{z^2}$ |
| | $d_{\bar{y}\bar{z}}$ | $e'_{g\mp}$ | $\frac{2}{\sqrt{6}}d_{xy} + \frac{2}{\sqrt{3}}d_{yz}$ |
| | $d_{\bar{x}\bar{z}}$ | | $\frac{2}{\sqrt{6}}d_{x^2-y^2} - \frac{2}{\sqrt{3}}d_{xz}$ |

TABLE S1. Relationships among the atomic $d$ orbitals in the cubic basis ($\bar{x}$, $\bar{y}$, $\bar{z}$) with $\bar{z}$ aligned to the along one of the octhaedral axis and trigonal basis ($x,y,z$) with $z$ axis aligned to (111) direction.

## III. BAND STRUCTURES AND DENSITY OF STATES

The band structure along the symmetry lines of the 2D hexagonal Brilloun Zone are reported in Fig.S2 for the different phases we discussed.
The corresponding projected density of states (PDOS) on in-plane and out-of-plane orbitals are shown in Fig. S3. The PDOS provides a quantitative evidence of the orbital occupation reported in the manuscript.

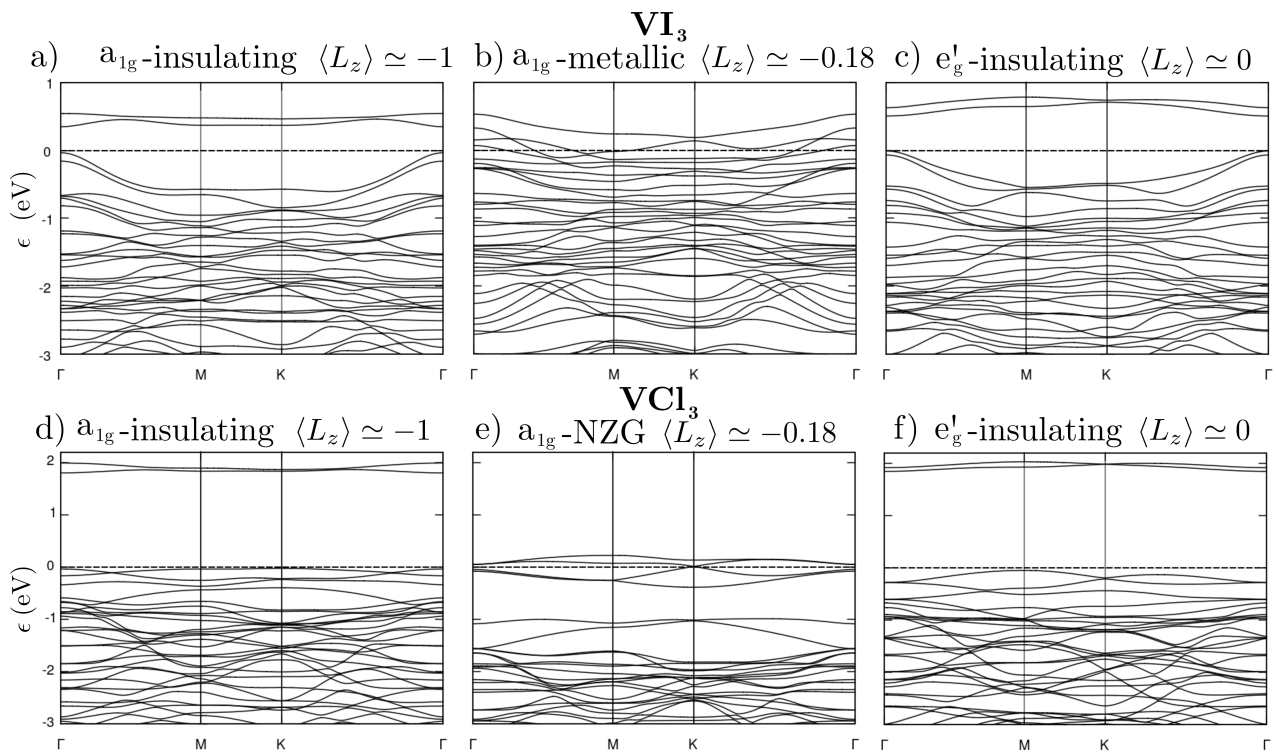

FIG. S2. Band structures of the different metastable states in a symmetric structure ($D_{3d}$), the zero of the energy is set to the valence band maximum for the insulating phases and the Fermi energy for metallic ones. a) VI$_3$ a$_{1g}$-metallic phase with partially quenched angular momentum, b) VI$_3$ a$_{1g}$-insulating phase with unquenched angular momentum, c) VI$_3$ e$_g$'-insulating phase with quenched angular momentum. d) VCl$_3$ a$_{1g}$-Near Zero Gap (NGZ) phase with partially quenched angular momentum; the NZG is opened by the SOC, e) VCl$_3$ a$_{1g}$-insulating phase with unquenched angular momentum, f) VCl$_3$ e$_g$' -insulating phase with quenched angular momentum.

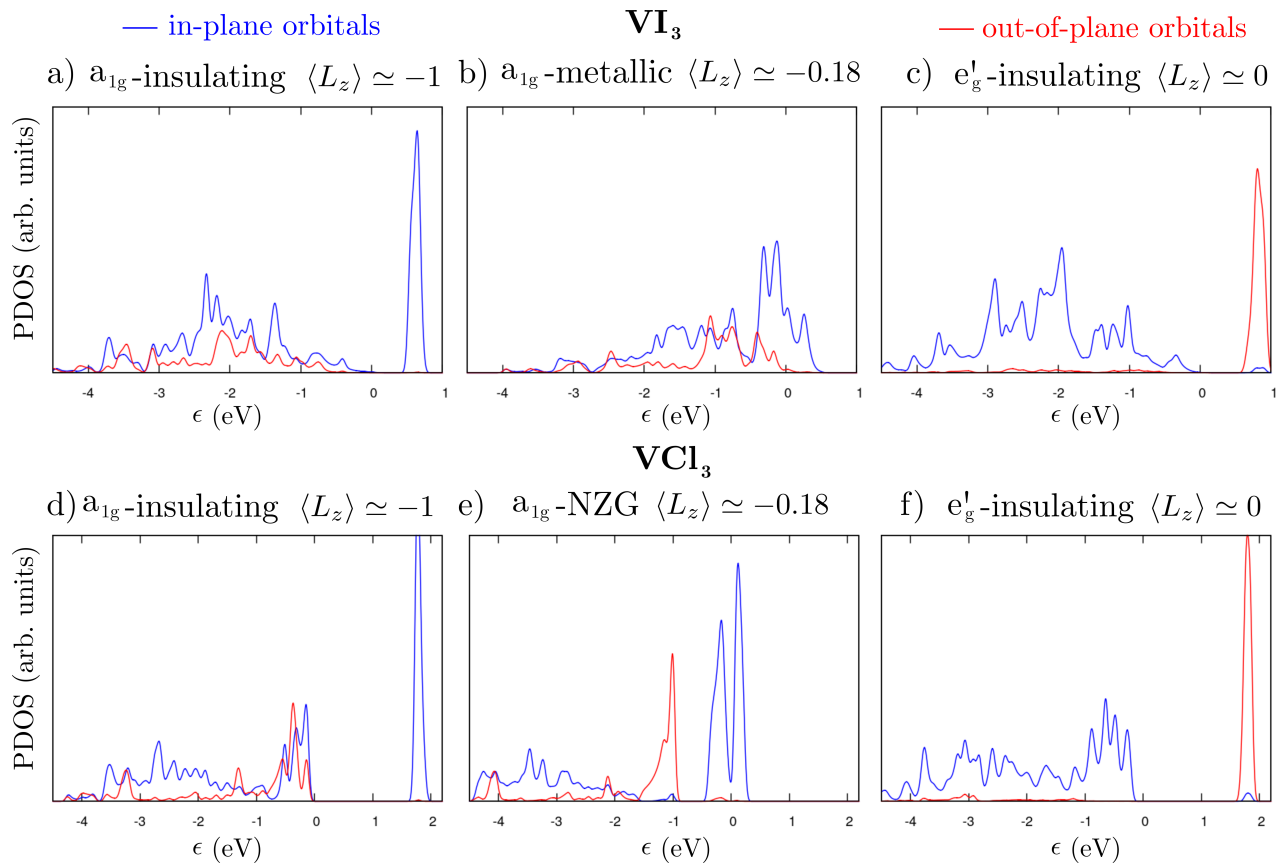

FIG. S3. Projected density of state on V-$d$ orbitals for all the studied phases. Out-of-plane orbital is proportional to the $a_{1g}$ state. The zero of the energy is set to the valence band maximum for insulating phases, the Fermi energy for metallic ones